\documentclass[aps,prb,twocolumn,showpacs]{revtex4-1}
\usepackage{bm}
\usepackage{graphicx}
\usepackage{amsmath,amsthm,amssymb}
\usepackage{dsfont}
\usepackage{soul}

\usepackage{color}

\begin{document}

\title{Geometric vector potentials from non-adiabatic spin dynamics}

\author{J. P. Baltan\'as,$^1$ H. Saarikoski,$^2$ A. A. Reynoso,$^{3,4}$ and D. Frustaglia$^1$}
\affiliation{$^1$Departamento de F\'isica Aplicada II, Universidad de Sevilla, E-41012 Sevilla, Spain}\email{baltanas@us.es}
\email{frustaglia@us.es}
\affiliation{$^2$RIKEN Center for Emergent Matter Science (CEMS), Saitama 351-0198, Japan}
\affiliation{$^3$Instituto Balseiro and Centro At\'omico Bariloche, Comisi\'on Nacional de Energ\'ia At\'omica, 8400 Bariloche, Argentina}
\affiliation{$^4$Consejo Nacional de Investigaciones Cient\'ificas y T\'ecnicas (CONICET), Argentina}

\date{\today}

\begin {abstract}

We propose a theoretical framework that captures the geometric vector potential emerging from the non-adiabatic spin dynamics of itinerant carriers subject to arbitrary magnetic textures. Our approach results in a series of constraints on the geometric potential and the non-adiabatic geometric phase associated with it. These constraints play a decisive role when studying the geometric spin phase gathered by conducting electrons in ring interferometers under the action of in-plane magnetic textures, allowing a simple characterization of the topological transition recently reported by Saarikoski {\it et al.} [Phys. Rev. B {\bf 91}, 241406(R) (2015)] in Ref.~\onlinecite{SVBNNF15}. 
 
\end{abstract}

\maketitle

\section{Introduction}

The role of geometric phases in diverse areas of physics and chemistry
has been the subject of intense research efforts since Berry's seminal
paper,\cite{berry} when non-trivial phases of geometrical origin
emerged as a widespread feature of Quantum
Mechanics. Among them, molecular physics is an important setting
where geometric phases play a crucial role.\cite{Mead1992} In
particular, a paradigmatic case occurs when the Born-Oppenheimer
approximation \cite{BO1927} is invoked. There, the nuclear coordinates
are considered to be slow when compared to the electronic degrees of
freedom, so that a treatment in which the electronic wave function
depends parametrically on the nuclear coordinates is appropriate. Mead
and Truhlar \cite {MT1979} showed that when the nuclear coordinates
encircle a closed path in the parameter space, the correct treatment
of the problem involves a term resembling a vector potential in the
effective Hamiltonian of the nuclear dynamics, which results in a
phase affecting the eigenfunctions. This phase, which in general
depends on the path described by the slow nuclear coordinates, was
eventually interpreted as an adiabatic geometric phase or Berry phase.

Inspired by these results, Aharonov {\em et al.} \cite{ABRPR}
considered a spin in the presence of a strong magnetic field in a
particular setup allowing for the Born-Oppenheimer
approximation. Rather than focusing their interest in the
  separation between fast and slow variables, they treated the kinetic
  term as an energy perturbation to the spin Hamiltonian. As a consequence of
  this purely algebraic approach, and without resorting to the intrinsic
  geometry of the problem, they found that the spin contribution was
   integrated into the kinetic terms in the form of a
  vector potential, just as expected within the Born-Oppenheimer
  approximation when an effective decoupling of charge and spin
  degrees of freedom is assumed. Later, Stern \cite{Stern} used this
  technique to study the effect of Berry phases in the conductance of
  spin carriers in 1D rings subject to magnetic textures (external
  magnetic fields of varying direction in space).

Here, we extend these ideas to the case of non-adiabatic spin dynamics where there is no clear separation between fast and slow degrees of freedom. We develop our theory by relaxing the adiabatic condition away from the perturbative regime considered by Aharonov {\em et al.}\cite{ABRPR} As a result we find expressions for non-adiabatic vector potentials and geometric phases, known as Aharonov-Anandan (AA) phases,\cite{AA87} satisfying a series of constraints. For illustration, we apply these findings to the problem of spin carriers confined in 1D conducting rings subject to in-plane field textures. This is partly motivated by our recent work \cite{SVBNNF15} on topological transitions in spin interferometers,
where non-adiabatic spin dynamics was proved to play a crucial role
near the transition point. The theory introduced here describes the reported topological transition in terms of an effective (adiabatic-like) Berry phase emerging from the actual non-adiabatic dynamics. 

The paper is organized as follows. In Section II we develop our general theory capturing geometric vector potentials and geometric phases in the case of non-adiabatic spin dynamics together with a series of constraints. In Section III, we apply this theory to the case of 1D rings subject to the action of in-plane topological field textures, where the constraints prove useful to identify topological features without the need to solve the full problem. We end with some concluding remarks summarizing the main results.

\section{Non-adiabatic spin dynamics in magnetic textures: general approach}

Already in his original paper, \cite{berry} Berry considered a spin
interacting with a magnetic field as an appropriate model to reveal
the presence of (adiabatic) geometric phases: a spin
state which adiabatically follows an external magnetic field describing a
closed trajectory in space accumulates a geometric phase factor
proportional to the solid angle subtended by the field. Let us recall
this system by considering the electronic Hamiltonian
\begin{equation}
{\mathcal H}=\frac{1}{2m}\boldsymbol{\Pi}^2+V({\mathbf r})+\mu{\mathbf B}(\mathbf r)\cdot\boldsymbol{\sigma},
\label{hamiltonian}
\end{equation}
where $\boldsymbol{\Pi}={\mathbf p}+(e/c){\mathbf A}_{\mathrm{m}}({\mathbf r})$,   
with ${\mathbf A}_{\mathrm{m}}({\mathbf r})$ the magnetic vector potential at position
${\mathbf r}$, $V({\mathbf r})$ an electrostatic potential
confining the electron motion, $\boldsymbol{\sigma}$ is the Pauli matrix vector
and ${\mathbf B}(\mathbf r)=B(\mathbf{r})\hat{\mathbf{n}}(\mathbf{r})$
is a magnetic field of varying magnitude and orientation, with $\hat{\mathbf{n}}(\mathbf{r})$
a unit vector defining its local direction. The field ${\bf B}({\bf r})$ may contain components from an external source [given by $\boldsymbol{\nabla}\times{\mathbf A}_{{\mathrm m}}({\mathbf r})$] together with effective components of dynamical origin as, e.g., an effective Rashba field arising from the spin-orbit coupling in the presence of an electric field.\cite{Rashba84} This particular case will be considered more explicitly in Sec. \ref{sec3}.

For an arbitrary ${\mathbf B(\mathbf r)}$, the eigenstates of the
Hamiltonian ${\mathcal H}$ defined in Eq.~(\ref{hamiltonian}) are
unknown. The approach followed in Ref.~\onlinecite{ABRPR} (see also
Ref.~\onlinecite{FR01}) starts by finding the local spin eigenstates
of the Zeeman term in Eq.~(\ref{hamiltonian}):
\begin{equation}
|\uparrow\rangle=\left(\begin{array}{c}\cos\frac{\alpha}{2}e^{-i\eta} \\ \sin\frac{\alpha}{2} \end{array}\right),\quad |\downarrow\rangle=\left(\begin{array}{c}-\sin\frac{\alpha}{2}e^{-i\eta} \\ \cos\frac{\alpha}{2} \end{array}\right),
\label{ad-spinors}
\end{equation}
which are locally (anti)aligned with the magnetic field's axis 
\begin{equation}
\hat{\mathbf{n}}(\mathbf{r})=\left(\sin\alpha(\mathbf{r})\cos\eta(\mathbf{r}),\sin\alpha(\mathbf{r})\sin\eta(\mathbf{r}),\cos\alpha(\mathbf{r})\right). 
\label{n}
\end{equation}
The states (\ref{ad-spinors}) coincide with the spin eigenstates of the full ${\mathcal H}$ only in the limit of adiabatic spin dynamics where the local Larmor frequency of spin precession, $\omega_{\rm s}=2\mu B(\bf{r})/\hbar$, is much larger than the frequency of orbital motion, $\omega_0 =v_{\rm F}/L$, with $v_{\rm F}$ the Fermi velocity and $L$ a representative length over which ${\bf B}({\bf r})$ changes direction. \cite{PFR03} 
The Hamiltonian can be written as the sum of a
diagonal and a non-diagonal projection onto the basis defined by the \emph{adiabatic} spin eigenstates, ${\mathcal H}={\mathcal H}_\mathrm{ ad}+{\mathcal H}_{\mathrm{nad}}$, respectively. The adiabatic
limit is achieved by taking ${\mathcal H}_{\mathrm{nad}} \rightarrow 0$, leaving ${\mathcal H}\approx {\mathcal H}_\mathrm{ ad}$. This procedure results in the identification of a geometric vector potential leading to Berry phases associated with the adiabatic nature of the spin dynamics. 

The adiabatic condition is guaranteed in Ref. \onlinecite{ABRPR} by treating the kinetic term as a perturbation to the Zeeman one in Eq. (\ref{hamiltonian}). However, we notice that this is only a sufficient condition and not a necessary one: indeed, the adiabatic regime can be achieved also in the opposite regime where the Zeeman energy is a perturbation to the kinetic one, as shown in Ref. \onlinecite{FR01}.\cite{note-1}

We extend the algebraic approach of Aharonov \emph{et al.}\cite{ABRPR} by considering the more general non-adiabatic case. According to the above discussion, this means that the kinetic energy must be at least of the same order of the Zeeman one. The non-adiabatic spin eigenstates of ${\mathcal H}$ can be rather complex, pointing along directions generally different from the one defined by the local magnetic field, $\hat {\mathbf n}({\mathbf r})$. In this situation, non-adiabatic AA geometric phases eventually emerge as a result of the intricate paths described by the spin eigenstates in the Bloch sphere. As a starting point, let us rewrite ${\mathcal H} = {\mathcal H}_\mathrm{d}+{\mathcal H}_{\mathrm{nd}}$ as the sum of diagonal (d) and non-diagonal (nd) projections onto the basis defined by the {\em non-adiabatic} spin eigenstates 
\begin{equation}
|+\rangle=\left(\begin{array}{c}\cos\frac{\theta}{2}e^{-i\delta} \\ \sin\frac{\theta}{2} \end{array}\right),\quad |-\rangle=\left(\begin{array}{c}-\sin\frac{\theta}{2}e^{-i\delta} \\ \cos\frac{\theta}{2} \end{array}\right),
\label{nad-spinors}
\end{equation}
locally quantized along the unit vector
\begin{equation}
\hat{\mathbf{l}}(\mathbf{r})=\left(\sin\theta(\mathbf{r})\cos\delta(\mathbf{r}),\sin\theta(\mathbf{r})\sin\delta(\mathbf{r}),\cos\theta(\mathbf{r})\right). 
\end{equation}
To this aim, we define projection operators on the corresponding subspaces given by
\begin{equation}
{\mathcal P}_{\pm}=\frac{1\pm\hat{\mathbf{l}}({\mathbf r})\cdot\boldsymbol{\sigma}}{2},
\end{equation}
We stress that $\hat {\mathbf l}({\mathbf r})$ generally differs from $\hat {\mathbf
  n}({\mathbf r})$ in the non-adiabatic regime. We further notice that, by the sole definition of eigenstates, it holds
\begin{eqnarray}
\mathcal{H}_{\mathrm d}&=&{\mathcal P}_{+}{\mathcal H}{\mathcal
  P}_{+}+{\mathcal P}_{-}{\mathcal H}{\mathcal P}_{-}\equiv\mathcal{H},\label{dH}\\
\mathcal{H}_{\mathrm{nd}}&=&{\mathcal H}-{\mathcal H}_{\mathrm d}={\mathcal
  P}_{+}{\mathcal H}{\mathcal P}_{-}+{\mathcal P}_{-}{\mathcal
  H}{\mathcal P}_{+} \equiv 0.\label{ndzero}
\end{eqnarray}
It requires some further elaboration to make the best of the formal expressions (\ref{dH}) and (\ref{ndzero}). Notice that $\boldsymbol{\Pi}$ does not commute with ${\mathcal P}_{\pm}$ (due to the presence of
${\mathbf p}=-i\hbar{\mathbf \nabla}$), and therefore mixes the spin subspaces. By following Refs.~\onlinecite{ABRPR} and \onlinecite{FR01}, we introduce an operator ${\mathbf A}$ responsible for the  ${\mathcal P}_{\pm}$-subspace mixing while $\boldsymbol{\Pi}-{\mathbf A}$ acts only within each subspace. This is accomplished without ambiguity by defining
\begin{equation}
{\mathbf A}=\boldsymbol{\Pi}-{\mathcal P}_{+}\boldsymbol{\Pi}{\mathcal P}_{+}-{\mathcal P}_{-}\boldsymbol{\Pi}{\mathcal P}_{-},
\end{equation}
which verifies $[\boldsymbol{\Pi}-{\mathbf A},{\mathcal P}_{\pm}]=0$ and
${\mathcal P}_{\pm}{\mathbf A}{\mathcal P}_{\pm}=0$. In Eq.~(\ref{hamiltonian}), by expanding 
\begin{equation}
\boldsymbol{\Pi}^2=(\boldsymbol{\Pi}-\mathbf{A}+\mathbf{A})^2=(\boldsymbol{\Pi}-\mathbf{A})^2+\mathbf{A}^2+ \{\boldsymbol{\Pi}-\mathbf{A},\mathbf{A}\},
\end{equation}
with $\{,\}$ the anticommutator, and considering
$[\mathbf{A}_{\mathrm{m}},\mathcal{P}_{\pm}]=0$ together with the
properties of projection operators
($\mathcal{P}^2_{\pm}=\mathcal{P}_{\pm}$ and
$\mathcal{P}_{+}+\mathcal{P}_{-}=\mathds{1}$), one arrives at 
\begin{equation}
\mathcal{H}_{\mathrm{d}}=\frac{1}{2m}\big[(\boldsymbol{\Pi}-\mathbf{A})^2+\mathbf{A}^2\big]+V(\mathbf{r})+\mu(\mathbf{B}\cdot\hat{\mathbf{l}})(\hat{\mathbf{l}}\cdot\boldsymbol{\sigma}),
\label{diagonal}
\end{equation}
and
\begin{equation}
\mathcal{H}_{\mathrm{nd}}=\frac{1}{2m}\big\{\boldsymbol{\Pi}-\mathbf{A},\mathbf{A}\big\}+\mu[\mathbf{B}\cdot\boldsymbol{\sigma}-(\mathbf{B}\cdot\hat{\mathbf{l}})(\hat{\mathbf{l}}\cdot\boldsymbol{\sigma})],
\label{nondiagonal}
\end{equation}
where we have dropped the dependence on $\mathbf{r}$ when convenient
for ease in notation. Moreover, the explicit evaluation of $\mathbf{A}$ gives
\begin{equation}
\mathbf{A}=\frac{i\hbar}{2}(\hat{\mathbf{l}}\cdot\boldsymbol{\sigma})\boldsymbol{\nabla}(\hat{\mathbf{l}}\cdot\boldsymbol{\sigma}).
\end{equation} 
Both $\mathcal{H}_{\mathrm{d}}$ and $\mathcal{H}_{\mathrm{nd}}$ are
written in the laboratory frame. We now turn to the non-adiabatic spin-eigenstate basis by introducing the local unitary operator
\begin{equation}
\mathcal{U}(\mathbf{r})=\left(\begin{array}{cc}\cos\frac{\theta}{2}e^{i\delta} & \sin\frac{\theta}{2}  \\ -\sin\frac{\theta}{2}e^{i\delta} & \cos\frac{\theta}{2} \end{array}\right),
\end{equation}
which diagonalizes $\mathcal{H}$ by acting on the non-adiabatic spinors (\ref{nad-spinors}) as 
\begin{equation}
\mathcal{U}(\mathbf{r})|+\rangle=\left(\begin{array}{c} 1 \\ 0 \end{array}\right),\quad\mathcal{U}(\mathbf{r})|-\rangle=\left(\begin{array}{c} 0 \\ 1 \end{array}\right).
\end{equation}
We first notice that 
\begin{equation}
\mathcal{U}\mathbf{A}\mathcal{U}^{\dagger}=\left(\begin{array}{cc} 0 & \mathbf{a}_{\rm g}^{+-}  \\ \mathbf{a}_{\rm g}^{-+} & 0 \end{array}\right),
\end{equation}
where  
\begin{equation}
\mathbf{a}^{s\bar{s}}_{\rm g} = \frac{\hbar}{2}\left(\sin \theta \boldsymbol{\nabla}\delta +i~s \boldsymbol{\nabla}\theta\right) 
\label{gmix}
\end{equation}
plays the role of a geometric mixing. Moreover, 
\begin{equation}
\mathcal{U}(\boldsymbol{\Pi}-\mathbf{A})\mathcal{U}^{\dagger}=\left(\begin{array}{cc} \boldsymbol{\Pi}-\mathbf{A}_{\mathrm{g}}^{+} & 0  \\ 0 &  \boldsymbol{\Pi}-\mathbf{A}_{\mathrm{g}}^{-} \end{array}\right),
\end{equation}
where
\begin{equation}
\mathbf{A}_{\mathrm{g}}^{s}=\frac{\hbar}{2}(1+s\cos\theta)\boldsymbol{\nabla}\delta
\label{Ag}
\end{equation}
is a geometric vector potential responsible for the AA geometric phases. Indeed, a direct computation proves the identity $\mathbf{A}_{\mathrm{g}}^{s}=i\hbar\langle s|\boldsymbol{\nabla}|s\rangle$, with $|s\rangle$ the non-adiabatic spin eigenstates of Eq.~(\ref{nad-spinors}). Thus, the AA geometric phase reads 
\begin{equation}
\phi^{s}_{\mathrm{g}}=\frac{1}{\hbar}\int\,\mathbf{A}_{\mathrm{g}}^{s}\cdot {\mathrm{d}\mathbf{r}} = i \int\,\langle s| \boldsymbol{\nabla}|s\rangle{\mathrm{d}\mathbf{r}}, 
\label{phi-geom-0}
\end{equation}
in agreement with the general expression introduced in Ref.~\onlinecite{AA87}.

Back to the Hamiltonian, we find
\begin{equation}
\mathcal{U}\mathcal{H}_{\mathrm{d}}\mathcal{U}^{\dagger}=\left(\begin{array}{cc}\mathcal{H}^{+} & 0  \\ 0 & \mathcal{H}^{-} \end{array}\right),
\end{equation}
where
\begin{equation}
\mathcal{H}^{s}=\frac{1}{2m}\big(\boldsymbol{\Pi}-\mathbf{A}_{\mathrm{g}}^{s}\big)^{2}+V_{\rm eff}^s
\label{diag_blocks}
\end{equation}
describes an electron gas corresponding to the non-adiabatic spin species $s$, with
\begin{equation}
V_{\rm eff}^s=\frac{1}{2m} \mathbf{a}^{s\bar{s}}_{\rm g} \cdot \mathbf{a}^{\bar{s}s}_{\rm g}+V(\mathbf{r})+s \ \mu \mathbf{B}(\mathbf{r})\cdot\hat{\mathbf{l}}(\mathbf{r}).
\label{Veff}
\end{equation}
The first term in Eq. \ (\ref{Veff}) represents a spin-independent geometric scalar potential acting as a local energy shift, which is typically negligible when compared to the kinetic energy in mesoscopic implementations.
We further notice that the Zeeman energy in (\ref{Veff}) can be expanded as 
$\mu B(\mathbf{r})\left(|\langle s|\uparrow \rangle|^2-|\langle s|\downarrow \rangle|^2\right)$. 
Similarly, we find 
\begin{equation}
\mathcal{U}\mathcal{H}_{\mathrm{nd}}\mathcal{U}^{\dagger}=\left(\begin{array}{cc} 0 & \mathcal{H}^{\pm}  \\ \mathcal{H}^{\mp} & 0 \end{array}\right),
\end{equation}
with
\begin{eqnarray}
\mathcal{H}^{s\bar{s}}&=&\frac{1}{2m} \left[ (\boldsymbol{\Pi}-\mathbf{A}_{\rm g}^s)\cdot \mathbf{a}_{\rm g}^{s\bar{s}}+\mathbf{a}_{\rm g}^{s\bar{s}} \cdot (\boldsymbol{\Pi}-\mathbf{A}_{\rm g}^{\bar{s}}) \right] \nonumber \\
&+& \mu B(\mathbf{r}) \ \left(\langle s|\uparrow \rangle \langle \uparrow|\bar{s} \rangle - \langle s|\downarrow \rangle \langle \downarrow|\bar{s} \rangle \right).
\label{nondiag_blocks}
\end{eqnarray}
Notice that the constraint imposed on
Eq. \ (\ref{nondiag_blocks}) by Eq.~(\ref{ndzero}) establishes a
definite link between $\mathbf{A}_{\rm g}^s$ and
$\mathbf{B}(\mathbf{r})$ that will be of particular importance in the
identification of the effective geometric phase introduced in
Ref.~\onlinecite{SVBNNF15}, as discussed below.

\section{Non-adiabatic spin dynamics in magnetic textures: 1D rings}
\label{sec3}

\begin{figure}
\includegraphics[width=\columnwidth]{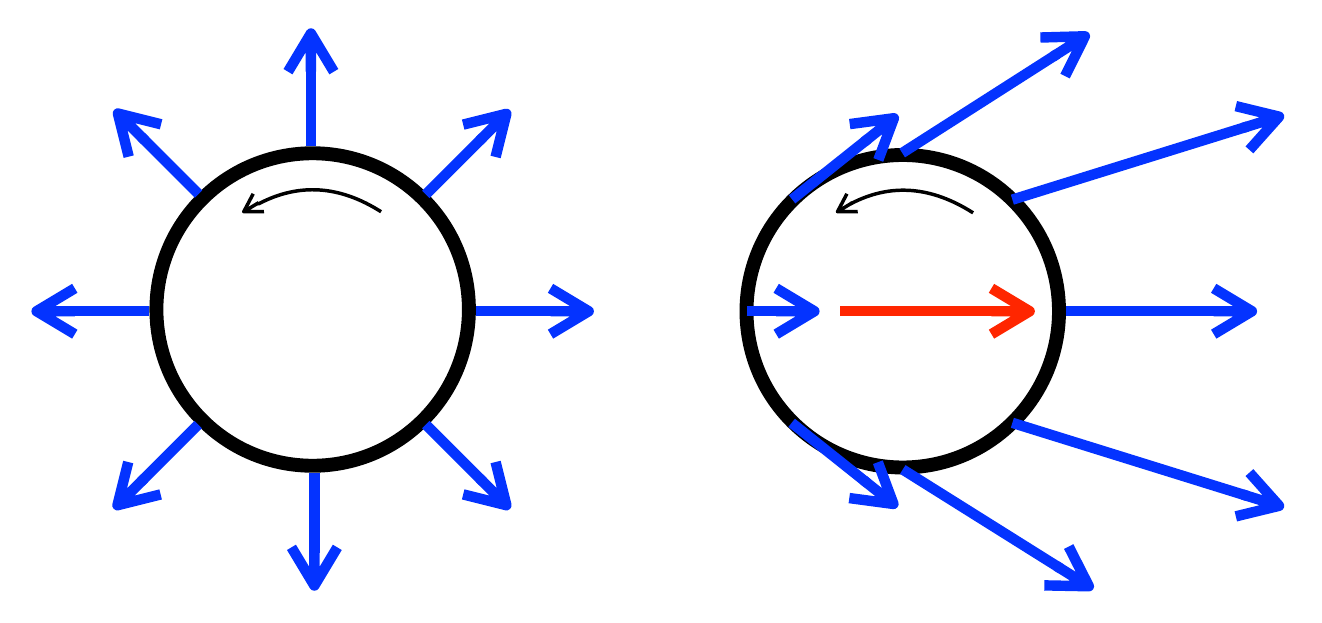}
\caption{Conducting ring with travelling spin carriers (thin arrow) subject to an hybrid magnetic-field texture: a radial effective field (originated internally by spin-orbit coupling) plus a uniform in-plane field (of external origin). The global field texture undergoes a topological transition from a rotating setup (left) to a waving one (right) as the uniform in-plane component increases.}
\label{fig-1}
\end{figure}

Among the several proposals for the manipulation of spin states by
guiding fields based on spin interferometry, one
by Lyanda-Geller stands out for its simplicity.\cite{lyanda-geller} There, he studied a 1D ring
interferometer subject to the combined action of internal (spin-orbit)
and external in-plane magnetic fields producing a magnetic texture
with variable topology: by tuning the magnitude of the external field
the global magnetic texture undergoes a topological transition, from a
rotating texture (enclosing the point of vanishing magnetic field in
the parameter space) to a waving one (the vanishing-field point is not
enclosed), see Fig.~\ref{fig-1}. By working within the limits of 
adiabatic spin dynamics, Lyanda-Geller concluded that the Berry phase 
accumulated by a spin state in a round trip would mirror the topological 
transition experienced by the magnetic texture by switching from $\pi$ to $0$, 
appearing as a topological imprint of the spin dynamics in the conductance of the ring.

Recently,\cite{SVBNNF15} we have pointed out that this description
turns out to be oversimplified: the spins are unable to follow the
magnetic field in the vicinity of the transition point since the
magnetic field vanishes and reverses its direction abruptly, which
casts serious doubts on the adiabatic character of the
dynamics. Despite this, we reported a phase dislocation in the
conductance as the remarkable signature of the topological transition
undergone by the magnetic field, close to what expected in the case
of adiabatic spin dynamics. This result is intriguing since, as noticed above, the
complexity of the non-adiabatic spin dynamics near the critical point
does not ease the way for an intuitive picture of the transition in
terms of geometric spin phases.

As we show below, the theoretical framework introduced in the previous
section provides a way to address the reported topological transition
in terms of an effective (adiabatic-like) Berry phase
emerging from the actual non-adiabatic spin dynamics. To this end, we
approach the particular case of electrons moving on a 1D ballistic
ring of radius $r$ and polar angle $\varphi$ in the presence on an
in-plane magnetic field texture [$\alpha=\pi/2$ in Eq. (\ref{n})],
generated from $\mathbf{A}_\mathrm{m}=A_{z}\hat{\mathbf{z}}$ with an
appropriate gauge choice (eventually, an additional component
$A_{\varphi}\hat{\boldsymbol{\varphi}}$ leading to an Aharonov-Bohm
flux could be considered). With the help of Eqs. (\ref{gmix}) and
(\ref{Veff}), Eq.~(\ref{diag_blocks}) reduces to
\begin{eqnarray}
\mathcal{H}^{s}&=&\frac{1}{2m}\left(\boldsymbol{\Pi}-\mathbf{A}_{\mathrm{g}}^{s}\right)^{2}+\frac{\hbar^{2}}{8mr^2}\left[\sin^{2}\theta\left(\frac{\partial\delta}{\partial\varphi}\right)^2+\left(\frac{\partial\theta}{\partial\varphi}\right)^{2}\right] \nonumber \\ 
&+&V(\mathbf{r})+s \mu B(\mathbf{r})\sin\theta\cos(\delta-\eta),
\label{diag-ring}
\end{eqnarray}
with
\begin{equation}
\mathbf{A}_{\mathrm{g}}^{s}=\frac{\hbar}{2r}(1+s\cos\theta)\frac{\partial\delta}{\partial\varphi}\hat{\boldsymbol{\varphi}}.
\label{Ag-ring}
\end{equation}
When the kinetic term in Eq. \ (\ref{diag-ring}) is dominant, the angular momentum of the moving charge is approximately conserved and the spatial part of the eigenfunctions takes the form $|\psi \rangle \sim e^{\pm i \ell\varphi}$ for counterclockwise (+) and clockwise (-) motion, with $\ell=k_{F}r$ and $k_{F}$ the Fermi wavevector. According to Eq. (\ref{phi-geom-0}), the AA geometric phase acquired by the spin $s$ carrier in a round trip is
\begin{eqnarray}
\phi^{s}_{\mathrm{g}}=\frac{1}{\hbar}\int\,\mathbf{A}_{\mathrm{g}}^{s}\cdot {\mathrm{d}\mathbf{l}}&=&\frac{1}{2}\int_{0}^{2\pi}(1+s\cos\theta)\frac{\partial\delta}{\partial\varphi}\mathrm{d}\varphi \nonumber \\
&=& n \pi + \frac{s}{2}\int_{0}^{2\pi}\cos\theta\frac{\partial\delta}{\partial\varphi}\mathrm{d}\varphi,
\label{phi-geom}
\end{eqnarray}
with ${\mathrm d}{\mathbf l}=r\mathrm{d}\varphi\hat{\boldsymbol{\varphi}}$ the elemental displacement along the ring and $n$ the winding (integer) number of the spin texture around the north pole of the Bloch sphere. The second term in Eq. \ (\ref{phi-geom}) is typically responsible for the fluctuations of the geometric phase appearing in complex spin textures.\cite{SVBNNF15} Likewise, the corresponding dynamical spin phase in a round trip is given by 
\begin{eqnarray}
\phi^{s}_{\mathrm{d}}&=&-\frac{m r^2}{\hbar^2 \ell}\int \left[ \mathcal{H}^{s} - \frac{1}{2m}\left(\boldsymbol{\Pi}-\mathbf{A}_{\mathrm{g}}^{s}\right)^{2} \right ] \mathrm{d}\varphi \nonumber \\
&=&-\int_{0}^{2\pi}\Big\{\frac{1}{8\ell}\Big[\sin^{2}\theta\Big(\frac{\partial\delta}{\partial\varphi}\Big)^2+\Big(\frac{\partial\theta}{\partial\varphi}\Big)^{2}\Big] \nonumber \\ 
&+& s\frac{mr^2}{\hbar^2 \ell}\mu B(\mathbf{r})\sin\theta\cos(\delta-\eta)\Big\}\,\mathrm{d}\varphi,
\label{phi-dyn}
\end{eqnarray}
after subtracting the kinetic contribution from $\mathcal{H}^{s}$ in Eq. (\ref{diag-ring}) and parametrizing the integral in terms of the polar angle of the ring. Notice that we have also dropped off the contribution from the confining potential $V(\mathbf{r})$ which essentially results in a constant phase shift. 

Moreover, the condition $\mathcal{H}^{s\bar{s}}=0$ imposed by Eq. (\ref{ndzero}) results in the
following equations for the real and imaginary parts of Eq. \ (\ref{nondiag_blocks}):
\begin{eqnarray}
&&\frac{\hbar^2}{4mr^2}\Big[\sin\theta\frac{\partial\delta}{\partial\varphi}\Big(\pm 2\ell-\frac{\partial\delta}{\partial\varphi}\Big)+s\frac{\partial^2\theta}{\partial\varphi^2}\Big] \nonumber \\
&+&\mu B(\mathbf{r})\cos\theta\cos(\delta-\eta)=0
\end{eqnarray}
and
\begin{eqnarray}
\frac{\hbar^2}{4mr^2}\Big[s\frac{\partial\theta}{\partial\varphi}\Big(\pm 2\ell-\frac{\partial\delta}{\partial\varphi}\Big)&-&\Big(\sin\theta\frac{\partial^2\delta}{\partial\varphi^2}+\cos\theta\frac{\partial\theta}{\partial\varphi}\frac{\partial\delta}{\partial\varphi}\Big)\Big] \nonumber \\
&+&\mu B(\mathbf{r})\sin(\delta-\eta)=0.
\end{eqnarray}
For simplicity, we focus on the case of counterclockwise spin carriers (with positive orbital quantum number $+\ell$) and take the semiclassical limit $\ell\gg 1$ corresponding to large
momentum or small Fermi wavelength, typical in mesoscopic rings.\cite{nagasawa2} By doing so we find that the previous expressions reduce to
\begin{eqnarray}
&&\frac{\hbar^2}{2mr^2}\sin\theta\frac{\partial\delta}{\partial\varphi}+\frac{\mu B(\mathbf{r})}{\ell}\cos\theta\cos(\delta-\eta)=0, \label{Re} \\
&s&\frac{\hbar^2}{2mr^2}\frac{\partial\theta}{\partial\varphi}+\frac{\mu B(\mathbf{r})}{\ell}\sin(\delta-\eta)=0,
\label{Im}
\end{eqnarray}
for magnetic-field strengths of, at least, the order of the ring's orbital-level spacing and/or containing a term proportional to the momentum as, e.g., in the case of effective, spin-orbit Rashba fields. 
The same approximation applies to the dynamical phase in Eq. (\ref{phi-dyn}) by neglecting the first terms under the integral sign. This implicitly assumes that all derivatives exist. Notice that this will be usually the case, with important exceptions as, e. g., spins passing over the poles of the Bloch sphere, where $\partial\delta / \partial\varphi$ diverges. In principle, our approach would not apply to those cases.

\subsection{Case study 1: AA phases in 1D Rashba rings}

In order to test the validity and soundness of the approach introduced above, we first apply it to the case of a 1D ring of radius $r$ subject to the sole action of Rashba spin-orbit
coupling displaying an effective radial field (see Fig. \ref{fig-1} left). This model has the advantage of being exactly solvable.\cite{FR04} The explicit solution shows that the
corresponding spin-eigenstates do not quantize along the direction of
the effective radial field but are lifted with a constant angle from the ring's plane. More precisely, Eq.~(\ref{nad-spinors}) reduces
to:
\begin{equation}
|+\rangle=\left(\begin{array}{c}\cos\frac{\theta}{2}e^{-i\varphi} \\ \sin\frac{\theta}{2} \end{array}\right),\quad |-\rangle=\left(\begin{array}{c}-\sin\frac{\theta}{2}e^{-i\varphi} \\ \cos\frac{\theta}{2} \end{array}\right),
\label{rashba-spinors}
\end{equation}
while the effective magnetic field reads ${\mathbf B}(\mathbf
r)=B_{\mathrm{R}}\left(\cos\varphi,\sin\varphi,0\right)$, with $\varphi$ is the polar angle on the ring's plane
and $B_{\mathrm{R}}$ the strength of the effective  Rashba field (momentum dependent and proportional to the orbital quantum number $\ell$). The tilt angle $\theta$ does not depend on $\varphi$ and is
given by $\tan\theta=\omega_{\mathrm{R}}/\omega_{\mathrm{0}}$, where
$\omega_{\mathrm{R}}= 2\mu B_{\rm R}/\hbar$ and $\omega_{\mathrm{0}}= \hbar \ell/m r^2$ are characteristic
Larmor and orbital frequencies, respectively (see Ref.~\onlinecite{FR04} for further details).

For the spinors (\ref{rashba-spinors}), Eqs.~(\ref{Ag-ring}), (\ref{Re}) and (\ref{Im}) reduce to
\begin{eqnarray}
\mathbf{A}_{\mathrm{g}}^{s}=\frac{\hbar}{2r}(1+s\cos\theta)&\hat{\boldsymbol{\varphi}}&,
\label{Ag-ring-Rashba} \\
\frac{\hbar^2}{2mr^2}\sin\theta+\frac{\mu B_\mathrm{R}}{\ell}\cos\theta&=&0, \label{Re-Rashba} \\
s\frac{\hbar^2}{2mr^2}\frac{\partial\theta}{\partial\varphi}&=&0,
\label{Im-Rashba}
\end{eqnarray}
respectively, where we have used $\delta=\eta=\varphi$ and
$\partial\delta/\partial\varphi=1$ due to azimuthal symmetry. It is straightforward
to see that Eqs.~(\ref{Re-Rashba}) and (\ref{Im-Rashba}) may be rewritten
as:
\begin{equation}
\tan\theta=\frac{2\mu B_{\mathrm{R}}/(\hbar)}{\ell\hbar/(m r^2)}=\frac{\omega_\mathrm{R}}{\omega_{0}},\qquad \frac{\partial\theta}{\partial\varphi}=0,
\end{equation}  
which simply means that the tilt angle of the corresponding
eigenspinors is constant and that its value depends on the 
adibaticity parameter $\omega_\mathrm{R}/\omega_{0}$ in the precise
manner reported in Ref.~\onlinecite{FR04} by direct calculation.

Moreover, an explicit calculation of the geometric phase from Eq. (\ref{phi-geom}) by using the
geometric vector potential (\ref{Ag-ring-Rashba}) gives
$\phi^{s}_{\mathrm{g}}=\pi(n+s\cos\theta)$ with $n=1$, which is exactly the AA geometric phase accumulated by a spin in a round trip (equal to half the solid angle subtended by the spin texture in the Bloch sphere). Again, this reproduces the result obtained in Ref.~\onlinecite{FR04}.

\subsection{Case study 2: topological transitions in 1D rings}

When a uniform in-plane field is considered in addition to the intrinsic Rashba spin-orbit contribution, the problem is no longer solvable by exact means. In Ref.~\onlinecite{SVBNNF15}, we 
showed that the total phase acquired in this situation by a spin carrier in a round
trip, $\phi^s=\phi^{s}_{\mathrm{d}}+\phi^{s}_{\mathrm{g}}$, undergoes a
transition determined by the topology of the total (Rashba plus
uniform) guiding field. In the following we identify the basics of
this transition by applying the approach introduced here.

From Eq. (\ref{Re}), the dynamical phase (\ref{phi-dyn}) can be written as  
\begin{equation}
\phi_{\rm d}^s = 
s\int_0^{2\pi} \frac{1}{2} \left( \frac{1}{\cos\theta}-\cos \theta \right) \frac{\partial \delta}{\partial \varphi} \ {\rm d}\varphi,
\label{phi-dyn-2}
\end{equation}
holding for $\ell \gg 1$. At first glance, here we recognize two contributions to $\phi_{\rm d}^s$: a fluctuating one proportional to $\cos \theta$ and a smooth one proportional to $1/\cos \theta$ [from Eq. (\ref{Re}) we see that the former does not diverge for vanishing $\cos \theta$]. In this way, by adding (\ref{phi-dyn-2}) to (\ref{phi-geom}) the total phase reduces to
\begin{equation}
\phi^s = s\int_0^{2\pi} \frac{1}{2\cos\theta}\frac{\partial \delta}{\partial \varphi} ~{\rm d}\varphi+ n \pi,
\label{phi-total}
\end{equation}
thanks to the cancelation of the terms proportional to $\cos
\theta$. The total spin phase (\ref{phi-total}) consists then of a
smooth dynamical contribution plus a topological one determined by the
parity of the winding number $n$, where $n\pi$ plays the role of and effective (adiabatic-like) Berry phase emerging from the non-adiabatic spin dynamics. A parity transition in $n$ would then explain
the results reported in Ref.~\onlinecite{SVBNNF15}. However, the actual 
existence of complex spin textures running over the poles of the Bloch sphere 
in the vicinity of the transition point results in the development of singularities in the terms proportional to $\partial \delta / \partial \varphi$ appearing in the geometric vector potential 
(\ref{Ag-ring}) and the total spin phase (\ref{phi-total}). This complicates the analysis 
near the transition point and a full picture remains so far incomplete (see, however, next paragraph). 

It is worthy of mention that, in the limit $\ell \gg 1$ considered here, the spin dynamics of the 
carriers maps into a time-dependent problem with localized spins 
subject to an external driving (by, basically, identifying the polar angle $\varphi$ with the time $t$ 
in the ring's case).\cite{RBSVNF17} This eventually leads to the finding of a scalar analogue of the geometric vector potential encoding the AA geometric phases accumulated by the spin
due to the driving.\cite{RBSVNF17} Moreover, it has been shown that a parity
transition in the effective Berry phase also exists to a great approximation 
in this case. The mapping to a time-dependent problem has the significant advantage to clarify the limits of our approach in terms of spin resonances at the same time that it opens a door to a new class of resonance experiments for the study of topological transitions in spin and other two-level systems.\cite{RBSVNF17}

\section{Conclusions}

We introduced an algebraic technique providing a closed expression of geometric vector potentials and geometric phases for spin carriers subject to arbitrary magnetic textures in the general case of non-adiabatic spin dynamics. More importantly, the theory imposes some dynamical constraints of particular importance in practice by allowing the identification of geometric and topological features without solving the full problem.
The work is based on previous developments on the perturbative induction of geometric vector potentials in the limit of adiabatic spin dynamics.\cite{ABRPR} We relaxed the adiabatic condition away from the perturbative regime. 

We illustrate the potentials of our approach by discussing two examples. We first reproduced the exact results\cite{FR04} of an analytically solvable problem on AA geometric phases in 1D Rashba rings. Secondly, we consider the more difficult problem of a conducting 1D ring subject to the combined action of in-plane Rashba and uniform fields.\cite{SVBNNF15} There we identify an effective Berry phase underlying the non-adiabatic dynamics as a key to single out the topological imprints left by the field texture.

Finally, we notice that the scope of our approach is best understood in the semiclassical limit of large momentum (where a dynamical decoupling emerges between charge and spin dynamics) by mapping the spin carrier problem into a time-dependent one with localized spins.\cite{RBSVNF17}

\begin{acknowledgments}
We thank J. Nitta and J. E. V\'azquez-Lozano for valuable comments and discussions. This work was supported by Project No. FIS2014-53385-P (MINECO, Spain) with FEDER funds and by Grants-in-Aid for Scientific Research (C) No. 17K05510  (Japan Society for the Promotion of Science). 
\end{acknowledgments}



\begin{thebibliography}{99}
\bibitem{note-1}
The reason is that the orbital frequency, $\omega_0$, is proportional only to the square root of the kinetic energy while the Larmor frequency, $\omega_{\rm s}$, is proportional to the Zeeman energy, instead.
\end{thebibliography}
\end{document}